\documentclass[pdflatex,sn-mathphys-num,iicol]{sn-jnl}

\usepackage{graphicx}%
\usepackage{multirow}%
\usepackage{amsmath,amssymb,amsfonts}%
\usepackage{amsthm}%
\usepackage{mathrsfs}%
\usepackage[title]{appendix}%
\usepackage{xcolor}%
\usepackage{textcomp}%
\usepackage{manyfoot}%
\usepackage{booktabs}%
\usepackage{algorithm}%
\usepackage{algorithmicx}%
\usepackage{algpseudocode}%
\usepackage{listings}%

\theoremstyle{thmstyleone}%
\newcommand{\ms}{M_{\odot}}
\newcommand{\lp}{\lambda_\Phi}
\usepackage{graphicx}
\usepackage{amsmath}
\usepackage{amssymb}
\usepackage{subfig}
\usepackage{booktabs}

\begin{document}

\title[White dwarfs in minimal dilatonic gravity]{White dwarfs in minimal dilatonic gravity}


\author*[1,2]{{Denitsa} \sur{Staicova}}\email{dstaicova@inrne.bas.bg}

\affil*[1]{\orgname{Institute for Nuclear Research and Nuclear Energy, 
        Bulgarian Academy of Sciences}, \orgaddress{\street{72 Tzarigradsko Shaussee}, \city{Sofia}, \postcode{1784}, \country{Bulgaria}}}
        

\abstract{We study static, spherically symmetric white dwarfs in minimal dilatonic gravity (MDG) -- a Brans–Dicke theory with fixed coupling $ \alpha^2=1/3  $ and free Compton length $\lp$. Solving the full relativistic structure equations we find that MDG white dwarfs are strictly sub-Chandrasekhar for every $  \lp $: the maximum mass falls from $  1.425\,M_\odot  $ in GR to $  1.27\,M_\odot  $ at $  200  $ km and $  1.09\,M_\odot  $ at $  500  $ km. The ratio by which MDG reduces the maximum mass is robust against the equation of state. Including rigid rotation near mass shedding, consistency with the most massive observed white dwarfs restricts $\lp \lesssim 300$ km. An independent gravitational-redshift bound gives $\lp\lesssim 720$ km. Because the dilaton mass is density-independent the same coupling produces a percent-level altitude dependence of the Kepler-inferred $  GM_\oplus  $, constraining it from the Solar-System side. To tackle the problem, we introduce a free-boundary collocation method with a linearized-exterior Robin condition that eliminates the exponential stiffness of shooting methods and gives access to the screened regime in double precision.}




\maketitle

\section{Introduction}
Compact objects -- white dwarfs, neutron stars and black holes -- are the end
states of stellar evolution and sit at the crossroads of astronomy, nuclear
physics and gravitation. White dwarfs are the least dense class and represent
the final stage of $\sim90\%$ of all stars. Their observational record extends
from William Herschel to the Sloan Digital Sky Survey \cite{SDSS} and to \emph{Gaia}: the DR3-based $100$\,pc samples now
contain $\sim8000$ white dwarfs with precise parallaxes and photometry
\cite{SDSS100pc}, and the mass--radius relation has been tested at the $6\%$
($1\sigma$) level using gravitational redshifts in common-proper-motion pairs,
independently of any theoretical $M(R)$ relation \cite{CPMpairs}. This means  white dwarfs now can serve as gravity probes.

White dwarfs are supported by the pressure of the degenerate electron gas, which
leads to the Chandrasekhar limit \cite{chandra}. A carbon--oxygen white
dwarf pushed over that limit by accretion explodes as a Type~Ia supernova, and
the interpretation of super-luminous events as super-Chandrasekhar progenitors
has motivated a large literature on modified-gravity white dwarfs: see
\cite{Kalita} for $f(R)$, and, more recently, $f(R,T,L_m)$ \cite{fRTLm}, 4D
Einstein--Gauss--Bonnet gravity \cite{4DEGB}, and scalar--tensor gravity with
crystallized interiors \cite{crystal}. Screening mechanisms have been studied
for white dwarfs in the Newtonian approximation in \cite{screening}, with the
general conclusion that screened scalars \emph{reduce} masses and radii and
never push the mass--radius curve above the Newtonian one. White dwarfs have
also been proposed as critical tests able to exclude entire classes of
modified-gravity models \cite{JainKouvaris}.

Here we consider white dwarfs in \emph{minimal dilatonic gravity} (MDG). MDG was
first proposed by O'Hanlon \cite{5th} to describe a fifth force of finite range. The model was developed extensively by Fiziev as an alternative description of
dark energy and dark matter \cite{fiziev1-1000,fiziev2013}, and is a
Brans--Dicke theory with $\omega=0$ and a nontrivial cosmological potential
$U(\Phi)$, locally equivalent to $f(R)$ gravity with a withholding potential
\cite{fiziev2013}. The vanishing Brans--Dicke parameter fixes the scalar
coupling to $\alpha^{2}=1/(2\omega+3)=1/3$, so MDG is a
one-parameter theory. Since only the range $\lp$ can be varied, and not the strength, this makes the confrontation with data more constraining. 

Compact static stars in MDG were formulated in \cite{fiziev2014}, the phase-space
structure of the static equations analyzed in \cite{FizievPhase}, and neutron
stars studied with polytropic \cite{fiziev2014,FizievMarinov} and, recently,
realistic equations of state confronted with NICER and GW170817 data
\cite{NSMDG2026}. White dwarfs in MDG have not, to our knowledge, been studied.
They probe dilaton Compton lengths $\lp\sim10^{2}$--$10^{5}$\,km, a window
complementary to the neutron-star one.

We find that MDG white dwarfs are strictly sub-Chandrasekhar for every $\lp$ --the opposite of what the super-luminous SN~Ia literature requires -- and that the observed population of massive white dwarfs constrains $\lp$ directly through the reduction of the maximum mass, complementing the redshift-based mass–radius test. We also show that the same fixed coupling makes the theory visible in Earth-orbit data at comparable ranges. To handle the problem numerically, we introduce a free-boundary collocation method that removes the exponential stiffness which has long challenged scalar-tensor stellar-structure calculations.

The article is organized as follows. Section~\ref{sec:eqs} presents the MDG
structure equations, Sec.~\ref{sec:eos} the equations of state,
Sec.~\ref{sec:numerics} the numerical methods, Sec.~\ref{sec:results} the
results, Sec.~\ref{sec:solar} the Earth and solar-system constraints, 
Sec.~\ref{sec:discussion} the discussion and Sec.~\ref{sec:conclusion} the conclusion .

\section{The equations of MDG}
\label{sec:eqs}
The action of MDG is \cite{fiziev2013}
$$ A_{g,\Phi}=\frac{c}{2\kappa} \int d^4x \sqrt{|g|}\,(\Phi R-2\Lambda U(\Phi)),$$
where $\kappa=8\pi G_N/c^4$, $\Lambda$ is the cosmological constant, $\Phi\in(0,\infty)$ is the dilaton and $U(\Phi)$ the cosmological potential. The dilaton modifies the gravitational coupling, $G(\Phi)=G_N/\Phi$; GR with $\Lambda$ is recovered for $\Phi\equiv 1$, $U\equiv 1$. The field equations split into trace and trace-free parts,
\begin{equation}
\begin{aligned}
  &\Box\Phi+\tfrac{2}{3}\Lambda\big(\Phi U_{,\Phi}(\Phi)-2U(\Phi)\big)=\tfrac{\kappa}{3}T, \\
  &\Phi \hat{R}_\alpha^\beta=-\widehat{\nabla_\alpha\nabla^\beta} \Phi-\kappa\hat{T}_\alpha^\beta .
\end{aligned}
\end{equation}
The dilaton couples to matter only through geometry, which makes it a dark component, contributing to $\epsilon_{eff}$ without direct coupling to the fluid.  Following \cite{fiziev2014} we adopt the withholding potential
\begin{align}
&U(\Phi)=\Phi^2+\tfrac{3}{16}\,\mathfrak{p}^{-2}\left(\Phi-1/\Phi\right)^2,\qquad \notag\\
 &V'(\Phi)\equiv\tfrac{2}{3}\big(\Phi U_{,\Phi}-2U\big)=\frac{1}{2\mathfrak{p}^{2}}\left(1-\Phi^{-2}\right),
\label{eq:potential}
\end{align}
with $\mathfrak{p}=\sqrt{\Lambda}\,\lp$, where $\lp$ is the single new parameter
relevant at stellar scales. Equation~(\ref{eq:potential}) is a \emph{withholding}
potential in the sense of \cite{fiziev2013}: it confines $\Phi$ to $(0,\infty)$
and has a unique stationary point at the GR value $\Phi=1$. Linearizing the trace
equation about it, $\Phi=1+\delta$, gives
$\Box\delta+\Lambda V''(1)\delta=\tfrac{\kappa}{3}T$ with
$\Lambda V''(1)=\lp^{-2}$, so $\delta$ is a Klein--Gordon field of mass
$m_\Phi=\lp^{-1}$ ($\hbar=c=1$) and $\lp$ is its Compton length
\cite{FizievPhase}. It is thus the curvature of $V$ at its minimum that sets the
dilaton mass.

For the static, spherically symmetric metric
\begin{equation}
 ds^2=e^{\nu(r)}dt^2-e^{\lambda(r)}dr^2-r^2d\Omega^2
\end{equation}
and a perfect fluid $T^\mu_\nu=\mathrm{diag}(\epsilon,-p,-p,-p)$ (units $G_N=c=1$; see Appendix~\ref{app:units}), the structure equations are \cite{fiziev2014}
\begin{align}
 &\frac{dm}{dr}=4\pi r^2 \epsilon_{\rm eff}/\Phi, \label{eq:dm}\\
 &\frac{dp}{dr}=-\frac{p+\epsilon}{r}\,\frac{m+4\pi r^3 p_{\rm eff}/\Phi}{\Delta-2\pi r^3 p_\Phi/\Phi}, \label{eq:dp}\\
 &\frac{d\Phi}{dr}=-4\pi r^2\, p_{\Phi}/\Delta, \label{eq:dPhi}\\
 &\frac{dp_{\Phi}}{dr}=-\frac{p_{\Phi}}{r\Delta}\Big(3r-7m-\tfrac{2}{3}\Lambda r^3+4\pi r^3 \epsilon_{\rm eff}/\Phi\Big)-\frac{2}{r}\epsilon_\Phi, \label{eq:dpPhi}
\end{align}
supplemented by the dilaton and cosmological equations of state
\begin{align}
&\epsilon_\Phi=p-\tfrac{1}{3}\epsilon+\tfrac{\Lambda}{8\pi}V'(\Phi)+\tfrac{p_\Phi}{2}\Pi, \quad \notag \\
&\epsilon_\Lambda=\tfrac{\Lambda}{8\pi}\big(U(\Phi)-\Phi\big),\quad p_\Lambda=-\tfrac{\Lambda}{8\pi}\big(U(\Phi)-\tfrac{1}{3}\Phi\big),
\label{eq:deos}
\end{align}
where $\Delta=r-2m-\tfrac{1}{3}\Lambda r^3$, $\epsilon_{\rm eff}=\epsilon+\epsilon_{\Phi}+\epsilon_{\Lambda}$, $p_{\rm eff}=p+p_{\Phi}+p_{\Lambda}$ and $\Pi=\big(m+4\pi r^3 p_{\rm eff}/\Phi\big)/\big(\Delta-2\pi r^3 p_\Phi/\Phi\big)$. At stellar scales the pure $\Lambda$ terms are negligible ($\Lambda\sim10^{-42}$
in dimensionless units), but $\Lambda V'$ is not: by (\ref{eq:potential}) it is independent of $\Lambda$ altogether, $\Lambda V'=(1-\Phi^{-2})/2\lp^{2}$, and it is this combination that carries the dilaton mass.

The unknowns are $m(r), p(r), \Phi(r), p_{\Phi}(r)$ with central conditions
$$m(0)=0,\quad \Phi(0)=\Phi_c,\quad p(0)=p_c,$$
$$p_{\Phi}(0)=\frac{2}{3}\left(\frac{\epsilon(p_c)}{3}-p_c \right)-\frac{\Lambda}{12\pi} V'(\Phi_c).$$
At the stellar edge $p(r^{*})=0$ the interior solution provides
$m^{*},\Phi^{*},p_{\Phi}^{*}$, and in the exterior ($p=\epsilon=0$) one must solve the boundary-value problem for the dilatonic sphere (``disphere'') with $\Phi\to1$ at the de Sitter horizon. The central value $\Phi_c$ is therefore not free: it is an eigenvalue fixed by this outer condition, and the problem is a two-point boundary-value problem rather than an initial-value one. 

\section{The equation of state}
\label{sec:eos}
The observed white-dwarf masses range as $M\!\in\![0.17\ms, 1.33\ms]$ with a peak near $0.6\,\ms$ and radii of $0.008$--$0.02\,R_\odot$ \cite{MR,SDSS100pc}, corresponding to central density $\rho_c \in [10^{5}, 10^{10}] \mathrm{g\,cm^{-3}}$, with $\rho_c^{peak}=3\times 10^6$. We employ three equations of state (EOS) of increasing realism \cite{ShapiroTeukolsky1983}.

\emph{(i) Polytropes.} In the non-relativistic ($k_F\ll m_e$) and
ultra-relativistic ($k_F\gg m_e$) limits the degenerate electron gas reduces to
$p=K\epsilon^{\Gamma}$ with $\Gamma=1+1/n$ , where $\Gamma=5/3$ and $4/3$, correspond to non-relativistic and relativistic polytropic fluid respectively: 
$$p_{\rm nonrel}=K_{\rm nonrel}\,\epsilon^{5/3},\qquad p_{\rm rel}=K_{\rm rel}\,\epsilon^{4/3},$$
$$K_{\rm nonrel}=\frac{\hbar^2}{15\pi^2 m_e}\left( \frac{3\pi^2 Z}{A m_N c^2} \right)^{5/3},$$
$$K_{\rm rel}=\frac{\hbar c}{12\pi^2}\left( \frac{3\pi ^2 Z}{A m_N c^2} \right)^{4/3},$$
with $A/Z=2.15$, $\epsilon=\rho c^2$ and $K$ written in terms of the neutron mass $m_N$ rather than the atomic mass unit. These admit exact Lane--Emden solutions
and are used throughout to fix the accuracy of the numerical solvers. Note that the $\Gamma=4/3$ mass, which in GR is independent of $\rho_c$, varies in MDG because the finite-range dilaton interaction $\lp$ breaks the homology of the $n=3$ polytrope.

\emph{(ii) Chandrasekhar EOS} The full ideal degenerate electron gas parametrized by the Fermi momentum $x=p_F/m_ec$:
$$P=\frac{\pi m_e^4c^5}{3h^3}\Big[x(2x^2-3)\sqrt{1+x^2}+3\,\mathrm{asinh}\,x\Big]$$
where $\epsilon=\rho c^2,\ \ \rho=\frac{8\pi m_e^3c^3\mu_e m_u}{3h^3}\,x^3,$
with mean molecular weight per electron $\mu_e=2$.

\emph{(iii) Salpeter-corrected EOS} Chandrasekhar plus the Coulomb lattice correction
\cite{Salpeter}, $P_C=-\tfrac{3}{10}(4\pi/3)^{1/3}Z^{2/3}e^2 n_e^{4/3}$
(we use $Z=6$, i.e. carbon composition). This correction reduces the pressure at fixed density and therefore shifts the absolute white-dwarf mass--radius relation relative to the Chandrasekhar EOS.
The effect on the \emph{relative} MDG modification is considerably smaller. Over the range $\lp=700$--$1500 {\rm km}$, inclusion of the Salpeter correction changes $\Delta m^*/m^*$ by only $0.5$--$1.3$ percentage points, and in the direction of slightly larger deviations from GR. The resulting $\lp$ bound is therefore robust to this EOS refinement. The correction is nevertheless important for the \emph{absolute} mass--radius calibration, since it produces a non-negligible shift of the GR results. 

\section{Numerical methods}
\label{sec:numerics}
We use the dimensionless units of \cite{fiziev2014} (Appendix~\ref{app:units}): $R_0=13.68$\,km, $M_0=9.263\,\ms$, $\varepsilon_0=6.466\times 10^{36}\,\mathrm{erg\,cm^{-3}}$. All results below were obtained with a Python implementation \footnote{code to be released at \url{https://github.com/dstaicova/MDG_WD} after publication} 
using two independent methods.

\emph{Shooting method:} The interior IVP is integrated from a series expansion at the center, the exterior is integrated past the surface and $\Phi_c$ is found by bisection on the sign of the diverging Yukawa mode. The intrinsic limitation is that the divergent interior mode amplifies any error in $\Phi_c$ by $e^{R_*/\lp}$: in double precision the method fails for $R_*/\lp \gtrsim 20$ due to the stiffness of the equation. 

\emph{Free-boundary collocation:.} We solve the interior as a two-point BVP on $x=r/r^*\in[x_0,1]$ with $r^*$ a free parameter, imposing central regularity at $x_0$ through the $O(r^2)$ expansions, and replacing the exterior by its linearization: since $|\Phi^*-1|\ll1$ in all physical cases, the disphere field obeys the Yukawa equation and can be condensed into a Robin condition at the surface,
\begin{equation}
p_\Phi(r^*)=\frac{\Delta^*}{4\pi r^{*2}}\Big(\frac{1}{\lp}+\frac{1}{r^*}\Big)\big(\Phi^*-1\big).
\label{eq:robin}
\end{equation}
Because the diverging mode never enters the computation, the $e^{R_*/\lp}$ amplification disappears entirely and no multiprecision is needed. With variables rescaled to $O(1)$ and a surface cut at $p=10^{-10}p_c$ (the $p\propto(1-x)^{5/2}$ surface behavior otherwise defeats mesh refinement), scipy's \texttt{solve\_bvp} reproduces the shooting results to $10^{-6}$ relative accuracy at $\sim 20\times$ lower cost, and continuation in $\lp$ extends the reach to $R_*/\lp\simeq 35$, where the residuals hit the double-precision floor of the near-cancellation $\epsilon_\Phi\to p-\epsilon/3+\Lambda V'/8\pi\approx 0$ at the in-matter equilibrium. Beyond that depth the deviations from GR are exponentially small and the GR solution can be used. A useful analytic control is the in-matter equilibrium of the trace equation, $\Phi_{\rm eq}=[1-q]^{-1/2}$, $q=\tfrac{16\pi}{3}(\epsilon-3p)\lp^2$, which our deep-screening solutions approach as expected. 

\begin{figure*}
\centering
\includegraphics[width=0.95\textwidth]{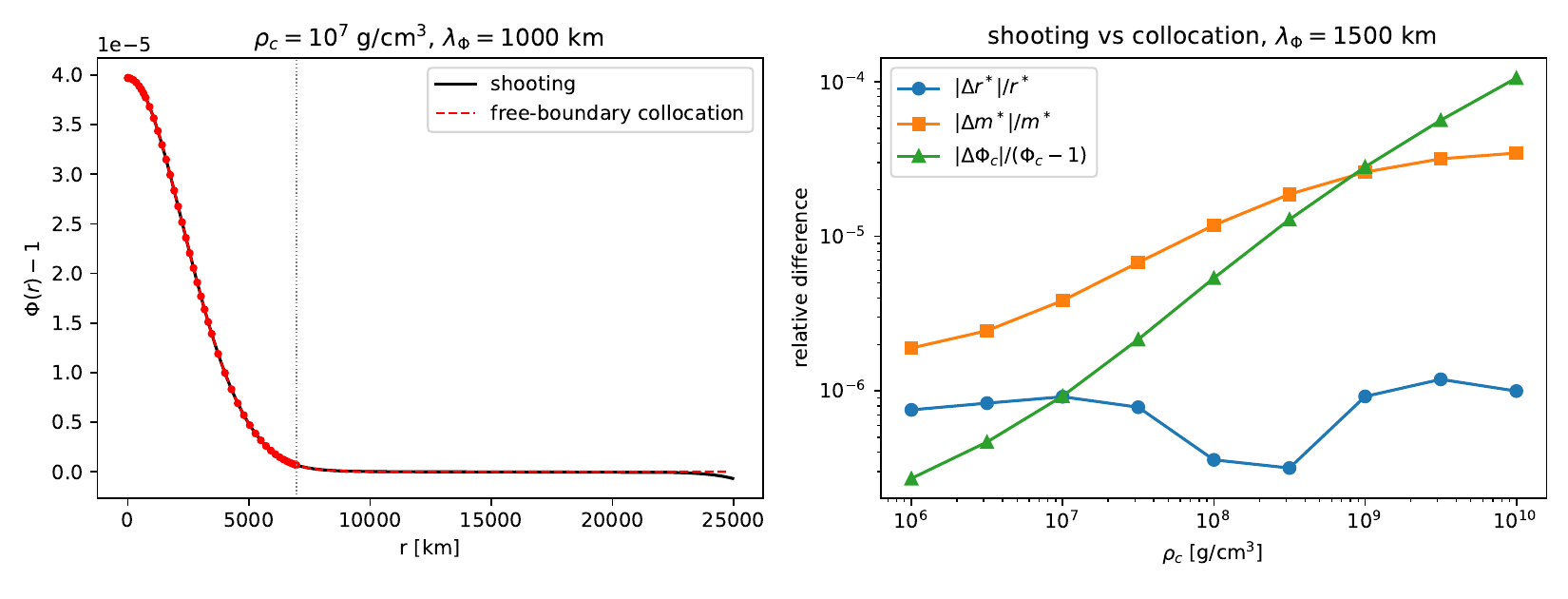}
\caption{Shooting method versus the free-boundary collocation method. Left: dilaton profile for a representative star; right: relative differences in $r^*$, $m^*$ and $\Phi_c-1$ along the $M(R)$ sequence for $\lp=1500$~km.}
\label{fig:methods}
\end{figure*}

The total gravitating mass $m_{\rm tot}$ (star $+$ disphere) is obtained by
integrating $dm/dr$ over the \emph{analytic} linearized exterior profile
$\Phi-1=(\Phi^*-1)(r^*/r)\,e^{-(r-r^*)/\lp}$, with $p_\Phi$ from its derivative.
The exterior equations admit both a decaying and a growing Yukawa mode, and any
error in the surface values excites the latter, so integrating outward directly
diverges. The analytic profile contains only the decaying mode, and the
integration is therefore stable for every $\lp$. Figure~\ref{fig:methods} demonstrates the agreement of the two independent solvers: the dilaton profiles coincide, and along an entire mass--radius sequence the stellar radii, masses and central dilaton values agree to the $10^{-7}$-$10^{-4}$ level with mild degradation toward high density.

\section{White-dwarf structure in MDG}
\label{sec:results}
This section presents the numerical results for white dwarfs in MDG: the structure of a fixed star across
the screening transition, the mass--radius relations and the resulting bound on $\lp$, and the effect of rotation.

\subsection{Solvers verification}
\label{sec:checks}
To test the two solvers described in Sec.~\ref{sec:numerics} we perform various checks on the accuracy of the calculation.

\emph{(i) The GR limit.} Setting $\lp\to0$ must return the TOV equations.
Table~\ref{tab:polytropes} compares the TOV limit with the exact Lane--Emden
solutions. The masses agree to
$\sim 10^{-4}$ and approach the Newtonian values as $\rho_c$ decreases: the residual is the physical
relativistic correction, proportional to the compactness $C = GM/Rc^{2}$,i.e. $|\Delta M|/M = a C$ with $a \simeq (6-8)$. For the Chandrasekhar equation of state the GR
maximum mass is $1.425\,\ms$ at $\rho_c\simeq3\times10^{10}\,\mathrm{g\,cm^{-3}}$. 

\emph{(ii) The screened limit.}  Deep inside a sufficiently screened star, the trace equation becomes approximately algebraic and drives the dilaton toward its in-matter equilibrium,
$\Phi_{\rm eq}=[1-q]^{-1/2}$, $q=\tfrac{16\pi}{3}(\epsilon-3p)\lp^{2}$. For the
$\rho_c=10^{7}\,\mathrm{g\,cm^{-3}}$ star the computed central value approaches
it monotonically, $(\Phi_c-1)/(\Phi_{\rm eq}-1)=0.46,\,0.77,\,0.91,\,0.97$ for
$R_*/\lp=4.7,\,10,\,18,\,35$ respectively. The convergence toward unity confirms that the numerical solutions recover the expected screened behaviour in the large-$R_*/\lp$ regime, including the range where the shooting method becomes unreliable. Additionally, the central field remains about 1 ($|\Phi_c-1|\lesssim10^{-4}$) throughout, ensuring that the linearization about $\Phi=1$ used to derive the Robin boundary condition (\ref{eq:robin}) is  justified.

\emph{(iii) The surface cut.} 
The stellar surface is formally defined by  $p(r^{*})=0$ but numerically we terminate the integration at $p=p_{\rm stop}p_c$. We have verified that this truncation changes the radius by only $\frac{\delta R}{R} \sim 10^{-4}$ 
 and the mass by $\frac{\delta M}{M} \sim 10^{-6}$
 for the white-dwarf sequences considered here. The mass converges faster because the low-pressure envelope contains negligible mass. Since the common truncation cancels in the MDG/GR ratio, we use  $p_{\rm stop}=10^{-10}$. In the polytrope checks of Table~\ref{tab:polytropes},
being absolute comparisons against analytic values, we use $p_{\rm stop}=10^{-14}$.

\subsection{Structure of the WD star and mass-radius relations}
\label{sec:transition}
\begin{figure*}
\centering
\includegraphics[width=0.95\textwidth]{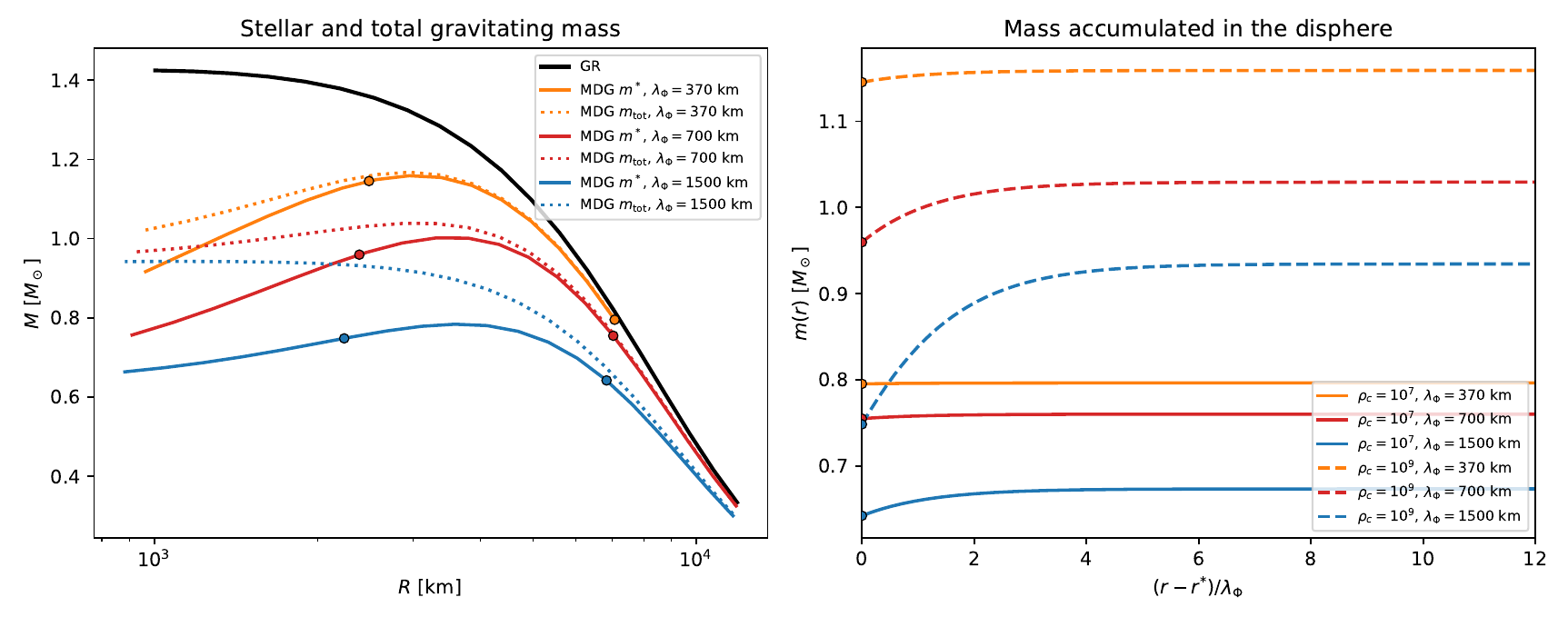}
\caption{Left: $M(R)$ in GR and MDG; solid $m^*$, dotted $m_{\rm tot}$
including the disphere. Right: gravitating mass accumulated outside the star, from the surface (circles) outward. Circles mark the stars for which we evaluate the disphere on the left.}
\label{fig:disphere}
\end{figure*}

On Figure~\ref{fig:disphere}, we present how the structure of a WD in the MDG theory varies with $\rho_c$ and $\lp$. Unless stated otherwise, we use Chandrasekhar EOS. 

For fixed central density $\rho_c=10^{7}\,\mathrm{g\,cm^{-3}}$, the results can be found in Table~\ref{tab:lam}. As $\lp$ is varied, the stellar mass falls monotonically with $\lp$ from within $0.7\%$ of $m_{\rm GR}$ at $\lp=200$\,km to a saturated $0.43\,m_{\rm GR}$ for $\lp\gg R_*$. 

If we keep fixed $\lp$, but vary $\rho_c$ (Table~\ref{tab:MR}), the behaviour is different: $m^*$ rises to a maximum and then \emph{decreases} toward high density, while $m_{\rm tot}$ continues to grow, so the star sheds mass into the disphere as it is compressed. At $\lp=700$\,km the disphere holds $0.7\%$ of the gravitating mass at $\rho_c=10^{7}\,\mathrm{g\,cm^{-3}}$ but $6.8\%$ at $10^{9}$; at $\lp=1500$\,km the same figures are $4.6\%$ and $19.9\%$. The right panel of Fig.~\ref{fig:disphere} shows that this mass accumulates over a few $\lp$ outside the surface and then saturates.

The saturated limit is analytic. For $\lp\to\infty$ the Yukawa term in
Eq.~(\ref{eq:newtlimit}) is unscreened everywhere and the Newtonian limit of MDG is simply gravity with $G_{\rm eff}=4G/3$, homology for a barotrope then gives $M\propto G^{-3/2}$, i.e.\ a mass ratio $(4/3)^{-3/2}=0.6495$. The computed total gravitating mass reaches $m_{\rm tot}/m_{\rm GR}=0.6499$ at $\lp=10^{5}$\,km. The light-field limit is therefore the strong-coupling limit of the theory, fixing the maximum MDG effect on a white dwarf.

The physical interpretation is that the dilaton (with fixed strength $  \alpha^2=1/3$) adds an extra attractive force within one Compton length. This means that there is positive dilaton pressure $ p_\Phi$ increasing the effective gravitating density and steepens the pressure gradient in Eqs.~(\ref{eq:dp})--(\ref{eq:dpPhi}). At fixed central density the star is therefore both smaller and lighter than in GR, with part of the “missing” mass stored in the exterior dilaton field (the disphere). The disphere recovers only a small fraction of this deficit: $\frac{(m_{\rm tot}-m^*)}{m_{\rm tot}}=0.02\%$, $0.26\%$, and $1.7\%$ for $\lp=200$, $500$, and $1000 {\rm km}$, respectively, reaching $\sim33\%$ only for $\lp\gtrsim10^4 {\rm km}$. Thus the disphere becomes a substantial fraction of the total mass only when $m^*$ has already fallen to roughly half its GR value.

If we compare the $M(R)$ relation when we vary both $\rho_c=10^{5.5}$--$10^{10.5}\,\mathrm{g\,cm^{-3}}$ and $\lp= 500-1500km$, we see that MDG white dwarfs are smaller and lighter than their GR counterparts at every central density, and the suppression grows with $\lp$ (from $3.1\%$ at $\lp=500\,$km to $18.9\%$ for $\lp=1500$\,km.) The deviation grows toward higher mass, reaching $9.3\%$ at $0.9\,\ms$ for $\lp=700$\,km.

\begin{figure*}
\centering
\includegraphics[width=0.95\textwidth]{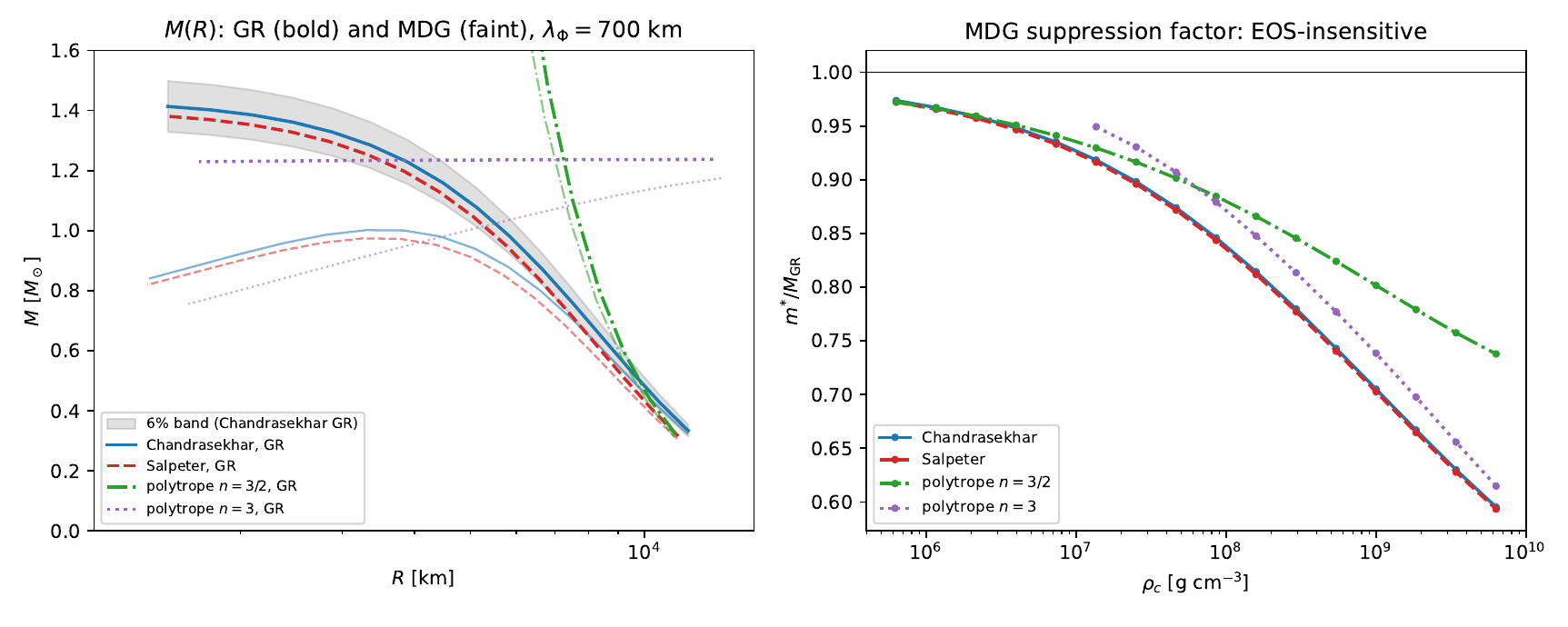}
\caption{Equation-of-state dependence at $\lp=700$\,km. Left: $M(R)$ in GR
(bold) and MDG (faint) for the four equations of state, with the $6\%$ band
around the GR Chandrasekhar curve. Right: the suppression factor
$m^{*}/M_{\rm GR}$}
\label{fig:eos}
\end{figure*}

Then we study the dependence of the results on the EOS of choice. This can be seen on Figure~\ref{fig:eos} (and the related Table~\ref{tab:eos}). By comparing the Chandrasekhar and Salpeter EOS, we see that the Coulomb correction lowers the absolute GR mass by $2$--$5\%$, but the
suppression factor $m^{*}/M_{\rm GR}$ changes by at most $0.26$ percentage points across the sequence. This means that the effect is insensitive to the microphysics and that any bounds on $\lp$ are not EOS dependent.

The maximum mass falls correspondingly (Table~\ref{tab:maxmass}): from
$1.425\,\ms$ in GR to $1.346,\,1.201,\,1.090$ and $0.996\,\ms$ for
$\lp=100,\,300,\,500$ and $700$\,km. MDG white dwarfs are therefore
strictly \emph{sub}-Chandrasekhar for every $\lp$, the opposite sign to the
$f(R)=R+\alpha R^{2}$ constructions invoked to explain super-Chandrasekhar
SNe~Ia \cite{Kalita}, and in the same direction as the Newtonian screening
result of \cite{screening}.

\begin{table}
\centering
\begin{tabular}{lccc}
\toprule
$\lp$ [km] & $m^{*}_{\rm max}$ [$\ms$] & $m_{\rm tot,max}$ [$\ms$] & $m^{*}_{\rm max}/m^{*,\rm GR}_{\rm max}$\\
\midrule
GR limit & 1.425 & 1.425 & 1.000\\
100 & 1.346 & 1.346 & 0.944\\
150 & 1.305 & 1.306 & 0.916\\
200 & 1.268 & 1.270 & 0.889\\
300 & 1.201 & 1.206 & 0.843\\
500 & 1.090 & 1.108 & 0.765\\
700 & 0.996 & 1.039 & 0.699\\
1\,000 & 0.887 & 0.976 & 0.623\\
1\,500 & 0.773 & 0.943 & 0.542\\
\bottomrule
\end{tabular}
\caption{Maximum stellar mass along the MDG sequences (Chandrasekhar EOS, $\mu_e=2$), where $m^{*}$ is the stellar mass and $m_{\rm tot}$ includes the disphere and is what a dynamical binary mass measures}
\label{tab:maxmass}
\end{table}

\begin{figure*}
\centering
\includegraphics[width=0.95\textwidth]{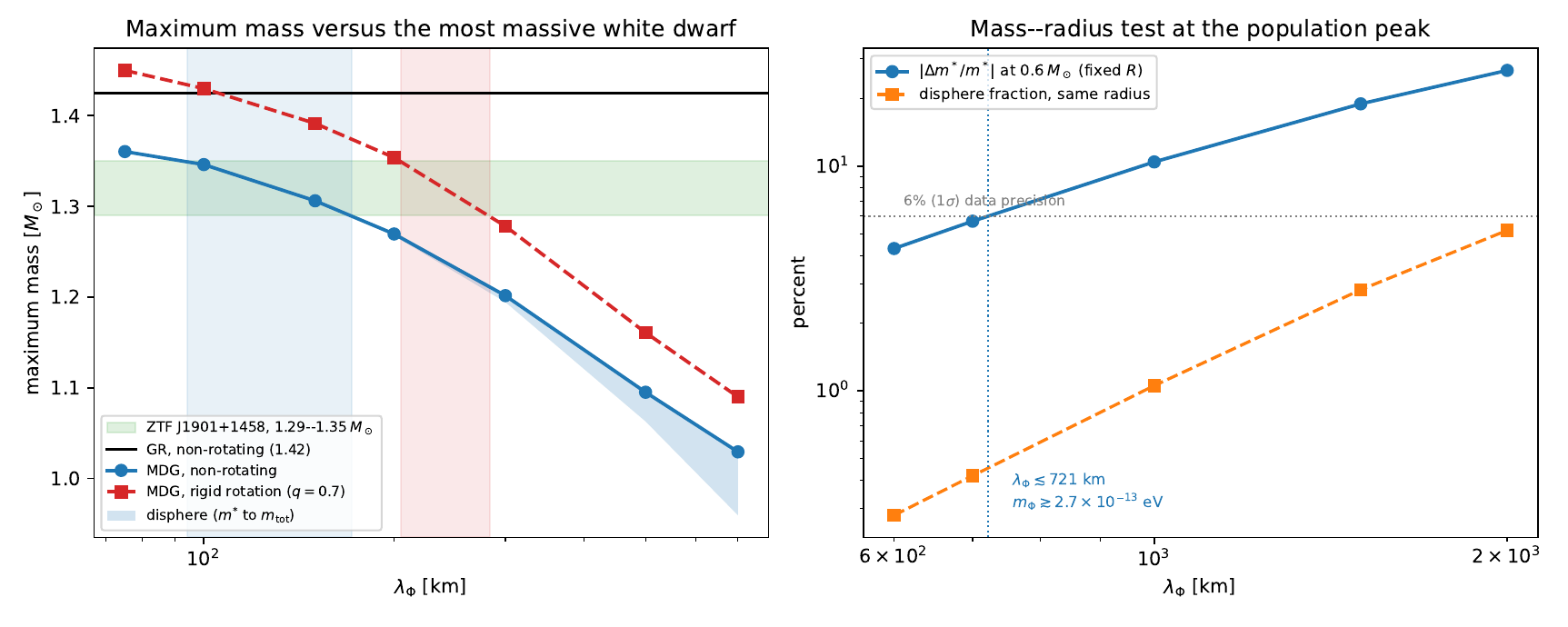}
\caption{Left: maximum gravitating mass $m_{\rm tot}$ against $\lp$ for
non-rotating and rigidly rotating WD. The shaded columns are the $\lp$ range corresponding to ZTF~J1901+1458. Right: deviation of the stellar mass from GR at fixed radius at the population peak, and the disphere fraction at the same radius, against $\lp$.}
\label{fig:constraint}
\end{figure*}

\subsection{Constraints from observed white dwarfs}
\label{sec:constraint}
MDG suppresses white-dwarf masses, so the observed population constrains $\lp$.
Our calculations use the Chandrasekhar and Salpeter EOS at $\mu_e=2$, which do not reproduce the composition, finite temperature and electrostatic corrections of any individual star. Therefore, the respective maximum GR masses, $1.425$ and $1.393\,\ms$, are the upper limits. The important quantity is not an absolute mass but the \emph{ratio} by which MDG reduces it, and that ratio is insensitive to the microphysics: the difference in $m^{*}_{\rm max}/m^{*,\rm GR}_{\rm max}$ between Chandrasekhar EOS and Salpeter EOS is about $0.2\%$ while the absolute maxima differ by $2.2\%$ (Table~\ref{tab:maxmass}, Table~\ref{tab:eos}).

The constraint therefore affects the maximal dynamical observed mass.  Whatever maximum mass a given composition
yields in GR, MDG reduces it by $5.7\%$, $11\%$ and $16\%$ at
$\lp=100,\,200$ and $300$\,km (if we do not take into account the disphere). Massive white dwarfs are observed close to the GR limit -- the most massive known, ZTF~J1901+1458, has
$M\simeq (1.29-1.35)\,\ms$ \cite{Caiazzo:2021xkk, SDSS100pc}. That puts a strong constraint on $\lp$, leading to $\lp$ of order $100$\,km for the static case. We do not use this as the strictest bound since we do not account for stellar compositions, or that the measured mass is not model-independent, relying on an assumed mass–radius relation.

If we take into account that the gravitational-redshift test of \cite{CPMpairs} constrains $m^{*}/R^{*}$, the relevant quantity is the deviation of the stellar mass at fixed radius. Requiring it
to remain within the observed $\lesssim6\%$ ($1\sigma$) agreement gives
$\lp\lesssim720$\,km, i.e.\
$m_\Phi\gtrsim2.7\times10^{-13}\,\mathrm{eV}/c^{2}$; at $2\sigma$,
$\lp\lesssim 1100$\,km. This bound shares the same limitation as the maximal mass one but the observational agreement is quoted differentially against a theoretical relation rather than against a single object, and the suppression factor is again EOS-insensitive (Table~\ref{tab:eos}).

Returning to ZTF~J1901+1458 ($ M\simeq1.3\,\ms$, $ R\simeq2.1  $-$ 2.6\times10^3$ km), the star sits close to the MDG mass limit. For $\lp=150$ km the non-rotating maximum is $1.305\,\ms$ (at $R\simeq2200$ km), compared with $1.425\,\ms$ in GR, so the observed mass range is reproduced for $\lp\simeq 100-170$ km. This concrete WD is not close to mass shedding ($P_{obs}=6.9$ min vs $\simeq1.5$\,s), so its own rotation does not enter. A star of the same mass, however, could be a rapid rotator -- the regime expected of accreting and merger-remnant progenitors -- and rigid rotation near mass shedding raises the MDG maximum to $1.43\,\ms$ at $\lp=100$\,km and $1.35\,\ms$ at $200$\,km (Fig.~\ref{fig:constraint}). Allowing for this we quote the more conservative $\lp\lesssim300$\,km. 

Figure~\ref{fig:constraint} also shows that at this bound the disphere holds $\sim0.5\%$ of the gravitating mass. The mass--radius signal and the disphere excess are controlled by the same coupling, and the former reaches the data precision an order of magnitude in $\lp$ before the latter becomes appreciable,
so they are not independent probes.

\subsection{Rotation}
\label{sec:rotation}
\begin{figure*}
\centering
\includegraphics[width=0.99\textwidth]{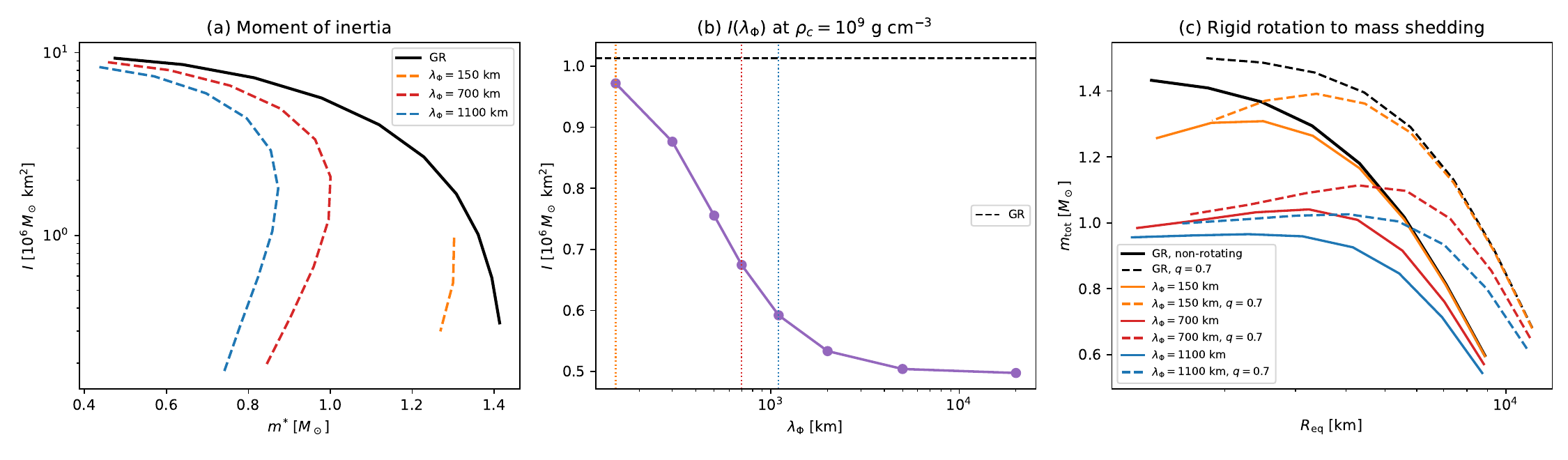}
\caption{Left: Moment of inertia $I$ against stellar mass along the $M(R)$. Middle: $I(\lp)$ for the $\rho_c=10^{7}\,\mathrm{g\,cm^{-3}}$ star, the dashed line is the GR value. Right: Rotating $M(R)$ relations (equatorial radius) from the Newtonian+Yukawa SCF code: non-rotating and rigidly rotating near-critical ($q=0.7$) sequences, GR limit versus MDG with $\lp=150, \lp=700$ and $1100$~km.}
\label{fig:rotMR}
\end{figure*}

Rotation is treated by two methods. The first is the relativistic Hartle expansion to linear order in the angular velocity $  \Omega  $, which yields the moment of inertia. The second is a non-perturbative Newtonian + Yukawa self-consistent-field (SCF) calculation that extends up to mass shedding. In the slow-rotation limit the two approaches overlap and agree: extrapolating the SCF sequence to vanishing rotational kinetic-to-potential energy ratio ($  T/W\to0  $) at $  \rho_c=10^7\,\mathrm{g\,cm^{-3}}  $ recovers the relativistic results to $  I_{\rm SCF}/I=1.0009  $ and $  M_{\rm SCF}/M=1.0007  $, consistent with the compactness $  GM/Rc^2\simeq1.7\times10^{-4}  $. Figure~\ref{fig:rotMR} shows the results. 

\emph{Slow rotation.} 
At linear order only the frame-dragging angular velocity $\omega(r)$ is excited. The scalar field and fluid perturbations appear only at $O(\Omega^2)$, consequently the mass and radius remain unchanged (they are even in $\Omega$)  while $\omega$ and $J$ are odd
\cite{Hartle,Staykov2014}. Defining $\bar\omega=\Omega-\omega$, the Einstein-frame
equation with metric functions $  \phi_E  $, $  \Lambda_E  $ and circumferential radius $  r_E  $ reduces to that of Ref.~\cite{Staykov2014} for $\omega_{BD}=0$:

\begin{equation}
\frac{d}{dr_E}\Big[e^{-(\phi_E+\Lambda_E)}r_E^4\frac{d\bar\omega}{dr_E}\Big]
= 16\pi\,(\epsilon+p)\,e^{\Lambda_E-\phi_E}\,r_E^4\,\bar\omega ,
\label{eq:framedrag}
\end{equation}

The total angular momentum $J$ is read from the exterior solution $\bar\omega=\Omega-2J/r^3$, giving the moment of inertia $I=J/\Omega$. In the GR limit the code recovers the exact Lane–Emden coefficient of the $n=3/2$ polytrope, $I/MR^2=0.2046$, to $3\times10^{-4}$.

For the $\rho_c = 10^9\,\mathrm{g\,cm^{-3}}$ star the moment of inertia is
suppressed across the screening transition: $I/I_{\rm GR} = 0.959,\ 0.865,\ 0.666,\ 0.584$ at $\lp = 150,\ 300,\ 700,\ 1100$\,km.The suppression is largely inherited from the smaller $  M^*  $ and $  R^*  $. The dimensionless shape factor $  k=I/(M^*R^{*2})  $ departs from its GR value by only a few percent in this range. Since a white dwarf cannot be identified without measuring $  M  $ and $  R  $, and $  I  $ is not measurable at comparable precision, the moment of inertia adds little constraining power. Its qualitative behaviour is nevertheless useful as a discriminator: MDG suppresses $  I  $, whereas $  R+aR^{2}  $ enhances it by $  30  $--$  40\%  $ for neutron stars \cite{Staykov2014}.

\emph{Rapid rotation.} Beyond slow rotation we use the Newtonian limit of MDG:
\begin{equation}
\Phi_{\rm grav}(\mathbf{x})=-G\!\int\! d^3x'\,\rho(\mathbf{x}')\,
\frac{1+\tfrac{1}{3}e^{-|\mathbf{x}-\mathbf{x}'|/\lp}}{|\mathbf{x}-\mathbf{x}'|},
\label{eq:newtlimit}
\end{equation}
 and compute rigidly rotating equilibria up to mass shedding with a Hachisu self-consistent-field method \cite{Hachisu}, the Yukawa
kernel expanded in modified spherical Bessel functions and the iteration
stabilized by holding $\rho_c$ and the axis ratio $q=r_{\rm pol}/r_{\rm eq}$
fixed while the length scale floats. The virial residual stays below
$5\times10^{-5}$ for all converged models, the non-rotating Newtonian limit
reproduces the Lane--Emden $n=3/2$ mass to $4\times10^{-6}$, and the SCF mass
matches the relativistic $m_{\rm tot}$ to $(1.5$--$5.7)\times10^{-4}$ across
$\lp=10^{3}$--$10^{5}$\,km. Note that the SCF mass is the total gravitating mass, because the Newtonian
Yukawa potential (\ref{eq:newtlimit}) already contains the exterior field energy, so there is no
Newtonian analogue of the split between $m^*$ and $m_{\rm tot}$.

Rigid rotation near mass shedding ($q=0.7$) raises the maximum mass by $\sim4$--$6\%$ in both theories: from $1.432$ to $1.499\,M_\odot$ in the Newtonian GR limit, from $1.041$ to $1.114\,M_\odot$ at $\lambda_\Phi=700$ km, and from $0.966$ to $1.026\,M_\odot$ at $\lambda_\Phi=1100$ km. Rotation therefore cannot mask the MDG suppression, nor the sub-Chandrasekhar prediction. Differentially rotating configurations, which in GR can support substantially super-Chandrasekhar masses, are left for future work.

\subsection{Comparison with other modified-gravity white dwarfs}
\label{sec:comparison}

Our results can be compared with several related modified-gravity studies. The Newtonian screening analysis of \cite{screening}, which considers chameleon, symmetron, and environment-dependent dilaton models, likewise finds that screened mass--radius curves do not exceed the Newtonian relation and that deviations increase toward lower density. MDG extends this analysis to fully relativistic stellar structure, includes the contribution of the exterior dilaton field to the total mass, and reduces the parameter space to the single range $\lp$ at fixed coupling $\alpha^2=1/3$. Perturbative $f(R)=R+\alpha R^2$ models, in contrast, admit both super- and sub-Chandrasekhar branches depending on the sign of the curvature correction, a feature that has been invoked to explain both super- and sub-luminous SNe~Ia \cite{Kalita}. MDG instead predicts a sub-Chandrasekhar shift for all $\lp$, making the two scenarios distinguishable at the bright end of the SN~Ia population. Recent $f(R,T,L_m)$ and 4D Einstein--Gauss--Bonnet models \cite{fRTLm,4DEGB} primarily modify $M(R)$ at high density, whereas the MDG deviation is largest at low-to-intermediate density and disappears in the screened limit. Finally, MDG suppresses the moment of inertia, while $R^2$ gravity enhances it \cite{Staykov2014}, which could be a useful discriminator between two $\omega_{\rm BD}=0$ theories that differ only in their scalar potential.

\section{Earth and solar-system tests}
\label{sec:solar}
In the weak field MDG is a massive Brans--Dicke theory with $\omega=0$, i.e.\ a
scalar of fixed coupling $\alpha^{2}=1/3$ and range $\lp$. The potential
Eq. (\ref{eq:potential}) provides no chameleon enhancement, since
$\Lambda V''\sim\lp^{-2}$ is density-independent.

\emph{Light deflection.} 
Light-deflection and Shapiro-delay tests are ambiguous for a massive dilaton. One common prescription absorbs the Yukawa into a redefined Newtonian potential and yields the position-dependent Eddington parameter \cite{Olmo,AlsingBertiWill}
$$\gamma(r)=\frac{3-e^{-r/\lambda_\Phi}}{3+e^{-r/\lambda_\Phi}}.$$
Applying the Cassini bound $ |\gamma-1|<2.3\times10^{-5} $ at $ r\simeq1 $ AU would give $\lambda_\Phi\lesssim 1.5\times10^7  $ km. Further discussions \cite{BerryGair, Clifton} , however, point that the experimental analyses assume a constant $ \gamma $ and if  the null geodesics are recomputed,  $\gamma=1$ identically. Therefore,  light-bending places no constraint on $  \lambda_\Phi$.

\emph{Perihelion precession.} An independent bound comes from the
periapsis advance, which the Yukawa enhances by $\delta\varpi=\pi\zeta$ per
orbit \cite{BerryGair}, with $\zeta\to0$ in both limits $\lp\to0$ and
$\lp\to\infty$, so each planet excludes a band in $\lp$. Using the Pitjeva residuals \cite{Pitjeva} for Mercury we recover $  \lp\lesssim1.9\times10^6  $ km on the short-range branch, already covering the entire white-dwarf window of interest.

\emph{Earth orbit.} The strongest and most directly relevant constraint for $  \lambda_\Phi\sim10^2  $–$  10^3  $ km is provided by Earth-orbit geodesy. Neither the Sun nor the Moon sources an appreciable dilaton field at satellite altitudes (both sit at $h/\lp\gtrsim100$, where $h=r-R$), while Earth satellite altitudes are comparable to $\lambda_\Phi$. Because the scalar is massive, Newton’s shell theorem fails: only a surface layer of thickness $\sim\lambda_\Phi$ contributes to the exterior field. For $R\gg\lp$ the surface is locally flat and the fractional excess in the effective $ GM$ inferred from a circular orbit is:
\begin{align}
&\delta(r)=\Big(1+\frac{r}{\lp}\Big) \times \notag \\
&\frac{(s-1)e^{-(r-R)/\lp}+(s+1)e^{-(r+R)/\lp}}{2s^{3}}\,
\frac{\rho_{s}}{\bar\rho},
\label{eq:delta}
\end{align}where $ s=\frac{R}{\lp} $, $ \rho_s\simeq3.3\,\mathrm{g\,cm^{-3}}  $ is the density of the crust/upper mantle and $\bar\rho$ is the the mean density of the Earth. 
For a uniform body $\delta(r)$ reduces at the surface to $\delta=\lp/2R$, but in our case, $\delta(r)$ is lower than that of 
uniform Earth. For $\lp=700$\,km, Earth's surface gravity would exceed its
Newtonian value by $\sim3\%$, the excess decaying with scale height $\lp$.

An orbit therefore returns $GM_{\rm eff}(r)=GM[1+\delta(r)]$, and satellites at
different altitudes disagree. The assumption-free statement is the predicted inconsistency between the $GM_\oplus$ inferred at GRACE and LAGEOS altitudes, $$\Delta=\delta(r_1)-\delta(r_2)
=
\begin{cases}
3.5\times10^{-5} & (\lambda_\Phi=100\,\mathrm{km}),\\
2.9\times10^{-3} & (\lambda_\Phi=300\,\mathrm{km}),\\
1.8\times10^{-2} & (\lambda_\Phi=700\,\mathrm{km}).
\end{cases}$$
Even allowing a conservative fractional precision $  \varepsilon\sim10^{-2}  $–$  10^{-3}  $ between the two determinations, $  \Delta  $ already exceeds $  \varepsilon  $ for $  \lambda_\Phi\gtrsim200  $–$  500  $ km. Because $  \Delta  $ rises steeply with $  \lambda_\Phi  $, the bound is only logarithmically sensitive to the precise value of $  \varepsilon  $ or $  \rho_s  $ and lies well below $  10^3  $ km.   We do not convert this into a formal bound: a
near-circular orbit absorbs a Yukawa entirely into the fitted $GM$, so the
signal lies in cross-mission comparison, and contemporary analyses share
GRACE-derived gravity models \cite{GGM05S}, whose mutual independence analysis is outside our scope. 

The Earth-orbit and maximum-mass bounds agree in placing $  \lambda_\Phi  $ at or below a few hundred kilometres. Stronger laboratory constraints exist, however: inverse-square-law tests with $  \alpha=1/3  $ (in particular the Eöt-Wash torsion-balance experiments) already require $  \lambda_\Phi\lesssim10^{-4}  $ m, or $  m_\Phi\gtrsim10^{-3}\,\mathrm{eV}/c^2  $ \cite{Adelberger} . The dilaton masses commonly adopted in the MDG stellar literature ($  m_\Phi\sim10^{-13}  $–$  10^{-11}\,\mathrm{eV}  $) are therefore in tension with local experiments and should be regarded as effective values appropriate only at low densities. We adopt them here for the same phenomenological reason -- to determine what white dwarfs would imply if the dilaton remained long-ranged, and thereby to close the window in which MDG could still have produced an observable astrophysical signature.

\section{Discussion}
\label{sec:discussion}

Our results show that MDG white dwarfs are smaller and lighter than their GR counterparts at every central density for every dilaton Compton length and no choice
of $\lp$ produces a super-Chandrasekhar branch. The theory therefore cannot
supply the massive progenitors that motivated much of the modified-gravity
white-dwarf literature.  

A distinctive feature of the model is that an MDG white
dwarf has not one but two masses. The disphere is gravitating but not baryonic, so a
distant observer measuring an orbit senses $m_{\rm tot}$, whereas the degenerate
matter available to burn, and the structure that determines the onset of
collapse, is $m^{*}$. A white dwarf would appear more massive dynamically
than the material it contains, and an accreting progenitor would reach its
structural limit at an apparent mass above it. This distinction is relevant to Type Ia supernovae, where the dynamical and burnable masses enter different parts of the progenitor argument: at fixed apparent mass, a smaller $m^*$
 implies less material available for 
$^{56}$Ni  production and hence a fainter event. Near the maximum-mass configuration the disphere fraction is only $  \sim0.03\%  $–$  0.6\%  $ for $  \lp \sim100$–$300 $ km and $6.8\% $ at $\lp =700$ km,  so the effect remains below current observational scatter for both SN Ia luminosities and dynamical-mass measurements.

\begin{figure*}
\centering
\includegraphics[width=0.95\textwidth]{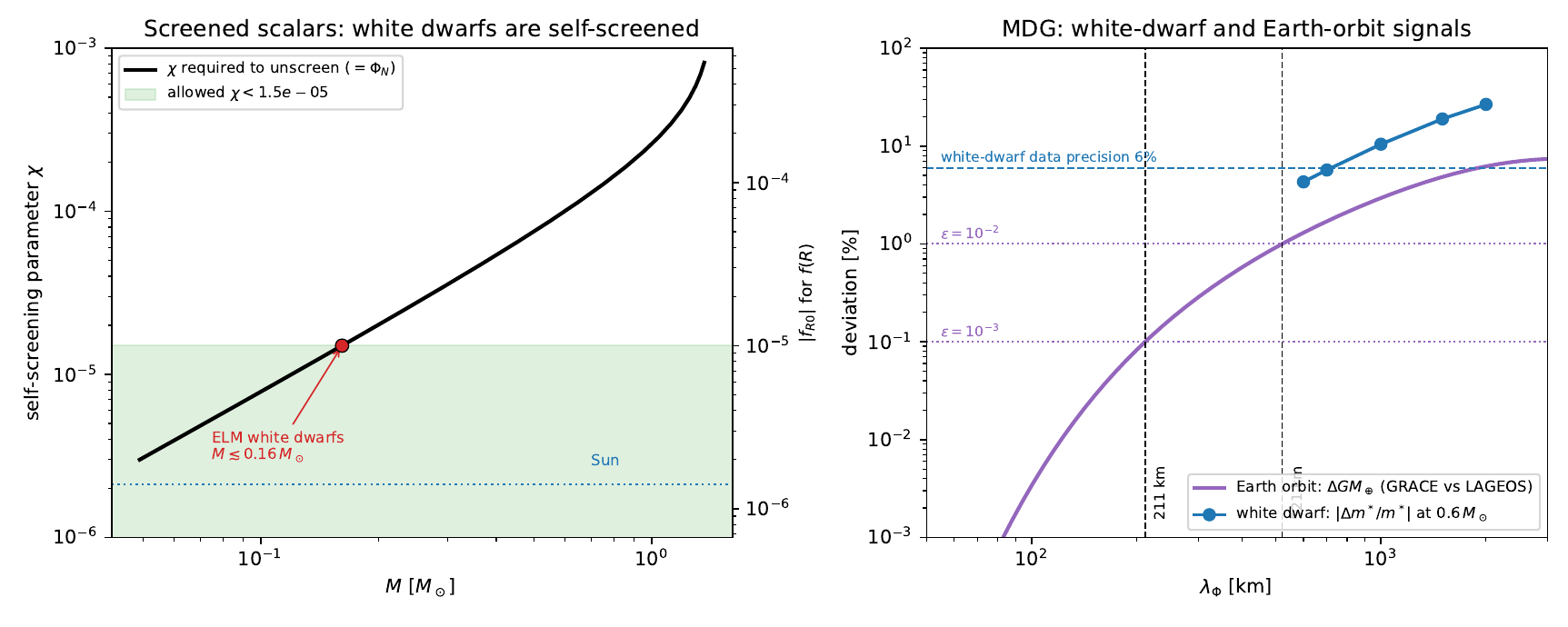}
\caption{Left: the self-screening parameter $\chi$ required to unscreen a star of given mass, the shaded band is the range allowed by astrophysical bounds, and the dotted line marks the Sun, with only extremely low mass (ELM) white dwarfs  remaining viable. Right: for MDG, the white-dwarf deviation at fixed radius and the predicted $GM_\oplus$ inconsistency between GRACE and LAGEOS altitudes, against $\lp$. }
\label{fig:pincer}
\end{figure*}

Figure~\ref{fig:pincer} summarizes the situation. Because $ \Lambda V''\sim\lambda_\Phi^{-2} $ is density-independent, the dilaton has the same Compton length in the Earth’s crust as in vacuum, and the fixed coupling $  \alpha^2=1/3$ cannot be weakened--only shortened in range. Consequently the white-dwarf and Earth-orbit signals are controlled by the same parameter $
\lp$ and probe the same few-hundred-kilometre scale through qualitatively different observables.

The problem is specific to the withholding potential rather than to $  \omega_{\rm BD}=0  $ gravity in general. In the Einstein frame the same conformal coupling $  \beta=1/\sqrt{6}  $ with a density-dependent mass produces chameleon screening. Along our sequences the Newtonian potential $  \Phi_N=GM/Rc^2  $ runs from $  7\times10^{-4}  $ at $  1.34\,M_\odot  $ to $  3\times10^{-5}  $ near $  0.3\,M_\odot  $. If we introduce a self-screening parameter $ \chi$ such that a star is unscreened only if $\Phi_N<\chi$ (for $f(R)$, $\chi=3/2|f_{R0}|$), ordinary white dwarfs are screened for every self-screening parameter $ \chi$ permitted by astrophysical bounds. Only extremely-low-mass objects ($  M\lesssim0.16\,M_\odot  $) cross the threshold. A dilaton light enough to alter white-dwarf structure is therefore heavy enough to be seen in Earth orbit, while one screened enough to evade Earth-orbit tests leaves white dwarfs unchanged. Theories screened by the Vainshtein mechanism evade this argument and lie outside the $  \omega_{\rm BD}=0  $ family considered here.

MDG was developed as a unified description of dark energy and dark matter \cite{fiziev1-1000,fiziev2013}. With $\lp\lesssim300$ km the exterior disphere contributes only a sub-percent correction to the total gravitating mass and the
bulk of the population lies within $1\%$ of its GR mass, so white-dwarf cosmochronology is unaffected. The maximum mass is more sensitive, being reduced by up to $16\%$, with rotation compensating only moderately ($4$--$6\%$ at maximum rotation), so single-degenerate progenitors would ignite at lower mass.
Evaluating this effect across the population is outside our scope.

While the result is negative for MDG in its present form, the free-boundary collocation scheme with a linearized-exterior Robin condition removes the exponential stiffness of massive scalar stellar-structure equations and extends naturally to screened scalar--tensor models. A relativistic treatment of the extremely low-mass, partially unscreened white dwarfs identified above is the most direct continuation.

\section{Conclusions}
\label{sec:conclusion}
We have presented a fully relativistic study of white dwarfs in minimal dilatonic gravity (MDG), using Chandrasekhar and Salpeter-corrected equations of state and two independent numerical methods -- the shooting method and the free boundary collocation method. The latter with a linearized-exterior Robin condition eliminates the $e^{R_*/\lp}$ error amplification that previously limited access to the screened regime. 

We find that MDG white dwarfs are strictly sub-Chandrasekhar for all $\lp$: the maximum mass decreases from $1.425\,\ms$ in GR to $1.346$, $1.268$ and
$1.090\,\ms$ for $\lp=100$, $200$ and $500$\,km -- equivalently, a reduction of $  5.7\%  $, $  11\%  $ and $  23.5\%  $ relative to the GR value for the same composition. The ratio is insensitive to the EOS (changing by only $  0.2\%  $ between Chandrasekhar and Salpeter), while the absolute maxima differ by $2.2\%$. Since the most massive observed white dwarfs lie close to the general-relativistic limit for their composition, consistency requires $ \lp \lesssim 300$km once rigid rotation is included. The gravitational-redshift mass–radius relation provides a weaker but better-controlled bound, $  \lp \lesssim 720$ km, i.e.\ $m_\Phi\gtrsim2.7\times10^{-13}\,\mathrm{eV}/c^{2}$.

Within the astrophysically allowed window the exterior disphere contributes only a sub-percent fraction of the total gravitating mass, the dimensionless shape factor differs from GR by a few percent, and rigid rotation near mass shedding raises the maximum mass by only $  \sim4\%$–$6\%$. None of these effects can yield a super-Chandrasekhar branch, but  the dimensionless shape factor discriminates MDG from $R + \alpha R^2$ theories.

Most importantly, the white-dwarf constraint is consistent with the Solar-System test. Because the dilaton mass is density-independent, the Kepler-inferred $GM_\oplus$ becomes altitude-dependent at the percent level for the same range of $  \lp$. The resulting limit of a few hundred kilometres provides an additional constraint on the same parameter range relying on entirely different observables.  

Taken together, the combined constraints restrict the model to $\lp\lesssim 300$ km  if it is to produce an appreciable modification of white-dwarf structure. A natural extension is therefore to introduce a density-dependent scalar mass, as in chameleon-like realizations such as the Hu--Sawicki and Starobinsky models, to which the numerical methods developed here transfer directly.

\section*{Acknowledgments}
This research was funded by Bulgarian National Science Fund grant number KP-06-N88/1.

\appendix
\section{Units}
\label{app:units}
Following \cite{fiziev2014}: $r\to R_0 r$, $m\to M_0 m$, $(p,\epsilon)\to\varepsilon_0(p,\epsilon)$ with
$$M_0\!=\!\Big(\!\frac{8\pi}{m_n}\!\Big)^2 M_P^3\!=\!9.263\,\ms, R_0\!=\!\frac{G_N}{c^2}M_0\!=\!13.68\,\mathrm{km},$$
$$\varepsilon_0=\frac{m_n^4c^5}{8\pi\hbar^3}=6.466\times10^{36}\,\frac{\mathrm{erg}}{\mathrm{cm}^3}, M_P=\sqrt{\hbar c/8\pi G_N},$$ and $\Lambda\to R_0^{-2}\Lambda$ ($\Lambda_{\rm obs}\simeq 1.9\times10^{-42}$ in these units).

\section{Tables}
\label{app:tables}

The following tables collect the numerical data underlying the figures and bounds of the main text.

\begin{table}
\centering
\begin{tabular}{ccccccc}
\toprule
$\rho_c$ [g\,cm$^{-3}$] & $R_{\rm GR}$ & $M_{\rm GR}$ & $R_{\rm LE}$ & $M_{\rm LE}$ & $R_{\rm MDG}$ & $M^*_{\rm MDG}$\\
\midrule
\multicolumn{7}{l}{\emph{relativistic, $\Gamma=4/3$ ($n=3$)}}\\
$1.328\times10^{6}$ & 28809 & 1.2378 & 28841 & 1.2383 & 30515 & 0.9699\\
$2.361\times10^{5}$ & 51232 & 1.2380 & 51288 & 1.2383 & 54480 & 1.0879\\
$4.199\times10^{4}$ & 91106 & 1.2381 & 91205 & 1.2383 & 94334 & 1.1687\\
\midrule
\multicolumn{7}{l}{\emph{non-relativistic, $\Gamma=5/3$ ($n=3/2$)}}\\
$1.666\times10^{6}$ & 9628 & 0.5227 & 9629 & 0.5229 & 8715 & 0.3049\\
$4.184\times10^{5}$ & 12122 & 0.2620 & 12122 & 0.2621 & 11124 & 0.1655\\
$1.051\times10^{5}$ & 15261 & 0.1313 & 15261 & 0.1313 & 14220 & 0.0902\\
\bottomrule
\end{tabular}
\caption{Polytropic EOS. Radii in km, masses in $\ms$; $A/Z=2.15$, surface cut $p_{\rm stop}=1\times10^{-14}\,p_c$. $R_{\rm LE},M_{\rm LE}$ are the exact Lane--Emden values built from the same $K$. The MDG columns use $\lambda_\Phi=5000$\,km.}
\label{tab:polytropes}
\end{table}

\begin{table}
\centering
\begin{tabular}{lcccc}
\toprule
$\lp$ [km] & $R_*/\lp$ & $r^{*}$ [km] & $m^{*}$ [$\ms$] & $m_{\rm tot}$ [$\ms$]\\
\midrule
GR limit & --- & 7082 & 0.8143 & 0.8143\\
200 & 35 & 7079 & 0.8085 & 0.8087\\
400 & 18 & 7067 & 0.7922 & 0.7934\\
500 & 14 & 7056 & 0.7811 & 0.7832\\
1\,000 & 7 & 6963 & 0.7117 & 0.7243\\
$1\times10^{4}$ & 0.62 & 6198 & 0.3796 & 0.5387\\
$1\times10^{5}$ & 0.061 & 6135 & 0.3532 & 0.5293\\
\bottomrule
\end{tabular}
\caption{White dwarf with $\rho_c=1\times10^{7}$\,g\,cm$^{-3}$ (Chandrasekhar EOS, $\mu_e=2$) as a function of $\lp$, $m_{\rm tot}$ includes the disphere. Rows with $R_*/\lp\gtrsim20$ ($\lp=200$\,km) use the collocation method}
\label{tab:lam}
\end{table}

\begin{table*}
\centering
\begin{footnotesize}
\begin{tabular}{r|rr|rrr|rrr}
\toprule
& \multicolumn{2}{c|}{GR} & \multicolumn{3}{c|}{MDG, $\lp=1500$ km} & \multicolumn{3}{c}{MDG, $\lp=5000$ km}\\
$\rho_c$ [g/cm$^3$] & $R$ & $M$ & $R$ & $M^*$ & $M_{\rm tot}$ & $R$ & $M^*$ & $M_{\rm tot}$\\
\midrule
$3.2\times10^{5}$ & 13428 & 0.250 & 13235 & 0.230 & 0.232 & 12411 & 0.164 & 0.188\\
$3.2\times10^{6}$ & 8862 & 0.597 & 8624 & 0.503 & 0.518 & 7996 & 0.338 & 0.421\\
$3.2\times10^{7}$ & 5579 & 1.017 & 5318 & 0.739 & 0.797 & 4942 & 0.509 & 0.685\\
$3.2\times10^{8}$ & 3295 & 1.291 & 3044 & 0.777 & 0.916 & 2885 & 0.597 & 0.850\\
$3.2\times10^{9}$ & 1814 & 1.400 & 1623 & 0.713 & 0.941 & 1579 & 0.623 & 0.917\\
$3.2\times10^{10}$ & 936 & 1.425 & 822 & 0.660 & 0.942 & 814 & 0.628 & 0.936\\
\bottomrule
\end{tabular}
\end{footnotesize}
\caption{Mass--radius sequences (radii in km, masses in $\ms$), Chandrasekhar EOS with $\mu_e=2$. $M^*$ is the stellar mass and $M_{\rm tot}$ includes the disphere.}
\label{tab:MR}
\end{table*}

\begin{table*}
\centering
\begin{footnotesize}
\begin{tabular}{r|rrr|rrr|r}
\toprule
& \multicolumn{3}{c|}{Chandrasekhar} & \multicolumn{3}{c|}{Salpeter} & \\
$\rho_c$ [g/cm$^3$] & $R$ & $M_{\rm GR}$ & $m^{*}/M_{\rm GR}$ & $R$ & $M_{\rm GR}$ & $m^{*}/M_{\rm GR}$ & $\Delta M_{\rm GR}$\\
\midrule
$1.0\times10^{6}$ & 10956 & 0.401 & 0.969 & 10581 & 0.381 & 0.967 & -5.0\%\\
$5.0\times10^{6}$ & 8115 & 0.683 & 0.944 & 7924 & 0.659 & 0.942 & -3.5\%\\
$2.5\times10^{7}$ & 5859 & 0.979 & 0.898 & 5756 & 0.951 & 0.896 & -2.9\%\\
$1.3\times10^{8}$ & 4101 & 1.204 & 0.827 & 4043 & 1.174 & 0.824 & -2.5\%\\
$6.3\times10^{8}$ & 2775 & 1.337 & 0.734 & 2742 & 1.306 & 0.731 & -2.4\%\\
$3.2\times10^{9}$ & 1814 & 1.400 & 0.634 & 1795 & 1.368 & 0.632 & -2.3\%\\
\bottomrule
\end{tabular}
\end{footnotesize}
\caption{Equation-of-state dependence at $\lambda_\Phi=700$\,km. Radii in km, masses in $\ms$. The Coulomb correction lowers the absolute GR mass by $2$--$5\%$ (last column), but the MDG suppression factor $m^{*}/M_{\rm GR}$ is nearly unchanged}
\label{tab:eos}
\end{table*}

\end{document}